\documentclass[aps,prd,onecolumn,groupedaddress,showpacs,nofootinbib,amssymb
]{revtex4}
\usepackage[dvips]{graphicx}
\usepackage{amssymb}
\usepackage{amsmath}
\usepackage{graphicx}
\usepackage{amsfonts}
\usepackage{bm}

\begin{document}

\title{Generalized $R^p$-attractor Cosmology in the Jordan and Einstein Frames: New Type of Attractors and Revisiting Standard Jordan Frame $R^p$ Inflation}
\author{S.D. Odintsov,$^{1,2,3}$\,\thanks{odintsov@ieec.uab.es}
V.K. Oikonomou,$^{4}$\,\thanks{v.k.oikonomou1979@gmail.com}}
\affiliation{$^{1)}$ ICREA, Passeig Luis Companys, 23, 08010 Barcelona, Spain\\
$^{2)}$ Institute of Space Sciences (ICE,CSIC) C. Can Magrans s/n,
08193 Barcelona, Spain\\
$^{3)}$ Institute of Space Sciences of Catalonia (IEEC),
Barcelona, Spain\\
$^{4)}$ Department of Physics, Aristotle University of
Thessaloniki, Thessaloniki 54124, Greece}

\begin{abstract}
In this work we shall study a new class of attractor models which
we shall call generalized $R^p$-attractor models. This class of
models is based on a generalization of the Einstein frame
potential of $R^p$ $f(R)$ gravity models in the Jordan frame. We
present the attractor properties of the corresponding
non-minimally coupled Jordan frame theory, and we calculate the
observational indices of inflation in the Einstein frame. As we
show, there is a large class of non-minimally coupled scalar
theories, with an arbitrary non-minimal coupling which satisfies
certain conditions, that yield the same Einstein frame potential,
this is why these models are characterized attractors. As we
demonstrate,  the generalized $R^p$-attractor models are viable
and well fitted within the Planck constraints. This includes the
subclass of the generalized $R^p$-attractor models, namely the
Einstein frame potential of $R^p$ inflation in the Jordan frame, a
feature also known in the literature. We also highlight an
important issue related to the $R^p$ inflation in the Jordan
frame, which is known to be non-viable. By conformal invariance,
the $R^p$ inflation model should also be viable in the Jordan
frame, which is not. We pinpoint the source of the problem using
two different approaches in the $f(R)$ gravity Jordan frame, and
as we demonstrate, the problem arises in the literature due to
some standard simplifications made for the sake of analyticity. We
demonstrate the correct way to analyze $R^p$ inflation in the
Jordan frame, using solely the slow-roll conditions.
\end{abstract}


\maketitle



\def\pp{{\, \mid \hskip -1.5mm =}}
\def\cL{\mathcal{L}}
\def\be{\begin{equation}}
\def\ee{\end{equation}}
\def\bea{\begin{eqnarray}}
\def\eea{\end{eqnarray}}
\def\tr{\mathrm{tr}\, }
\def\nn{\nonumber \\}
\def\e{\mathrm{e}}

\section{Introduction}

The physics of the primordial Universe is currently in the focus
of both theoretical and observational cosmologists. The next
fifteen years will be sensational for physicists since several
experiments which will probe the early Universe, will commence
their operation, like the stage 4 Cosmic Microwave Background
(CMB) experiments \cite{CMB-S4:2016ple,SimonsObservatory:2019qwx}
and the gravitational wave future interferometers like LISA
\cite{Baker:2019nia,Smith:2019wny}, DECIGO
\cite{Seto:2001qf,Kawamura:2020pcg}, BBO
\cite{Crowder:2005nr,Smith:2016jqs} and other experiments
\cite{Weltman:2018zrl,NANOGrav:2020bcs,NANOGrav:2020spf}. The main
aim of these experiments is to probe the early Universe in two
ways: firstly to observe whether a relic stochastic gravitational
wave background which was generated primordially exists in the
Universe, and secondly whether any tensor modes are imprinted in
the CMB, the so called $B$-modes (curl modes). Any observation of
either the aforementioned two, will verify the existence of the
inflationary era. Inflation
\cite{inflation1,inflation2,inflation3,inflation4} is one of the
most appealing candidate scenarios for describing the early
Universe, with alternatives being bouncing cosmologies
\cite{bounce0,bounce1,bounce2,bounce3,bounce4,bounce5,bounce6}.
The inflationary scenario will be put into test the next fifteen
years, so it is of profound important to study many aspects of
inflationary dynamics. Inflationary scenarios can arise from many
theoretical frameworks, with the two mainstreams descriptions
being scalar field theory
\cite{inflation1,inflation3,Baumann:2009ds} and modified gravity
\cite{reviews1,reviews2,reviews3}.

One appealing property of the scalar field models is that these
can be classified to attractor models \cite{alpha1}, and several
works already appear in the literature studying the attractor
properties of scalar models, see for example Refs.
\cite{alpha1,alpha2,alpha3,alpha4,alpha5,alpha6,alpha7,alpha7a,alpha8,alpha9,alpha10,alpha11,alpha12,alpha13,alpha14,alpha15,alpha16,alpha17,alpha18,alpha19,alpha20,alpha21,alpha22,alpha23,alpha24,alpha25,alpha26,alpha27,alpha28,alpha29,alpha30,alpha31,alpha32,alpha33,alpha34,alpha35,alpha36,alpha37,Ivanov:2021ily}
and references therein. In this paper the focus is on a new type
of attractor models which we shall call generalized
$R^p$-attractor models. These models are based on a scalar
potential which is a direct generalization of the Einstein frame
potential corresponding to the Jordan frame $R^p$ gravity of the
form $f(R)=R+\beta R^p$. We shall study the general features of
the generalized $R^p$-attractor models and specifically we shall
present the features of the non-minimally coupled scalar field
theory which corresponds to the Einstein frame $R^p$-attractor
models scalar potential. As we shall show, the generalized
$R^p$-attractor models in the Einstein frame generate a viable
phenomenology, including the Einstein frame potential
corresponding to the Jordan frame $R^p$ theory. This latter
feature has also been pointed out in the literature
\cite{Motohashi:2014tra,Renzi:2019ewp}. However, it is known from
the literature that power-law $R^p$ inflation in the Jordan frame
is not a viable scenario. This is in contrast to the general rule
that dictates that conformal invariant quantities of conformally
related theories in the Jordan and Einstein frames, should be the
same in the two frames \cite{kaizer,newsergei,Stabile:2013eha}. In
order to pinpoint the source of the problem, we present two
formalisms of Jordan frame $R^p$ inflation. As we demonstrate, the
two approaches lead to different results, and as we show, the
problem with the standard approach of $R^p$ inflation in the
literature is probably a simplification made for the sake of
analyticity for Jordan frame $R^p$ inflation. As a concluding
remark, our work indicates that although generalized $R^p$
inflation can be difficult to study in the Jordan frame, the
Einstein frame potential is rather easy to work with and the
conformal invariant quantities should be identical, thus the
Einstein frame study suffices to describe the generalized $R^p$
inflation. Our study could have potentially interesting
implications in neutron stars, since inflationary potentials and
inflationary modified gravity theories are frequently being used
in these contexts, see for example
\cite{Sotani:2017pfj,Astashenok:2021peo,Feola:2019zqg,Staykov:2014mwa,Stabile:2013eha}.

This work is organized as follows: In section II we present the
standard $R^p$ inflation scenario in the Einstein frame and we
introduce the generalized $R^p$-attractor potential in the
Einstein frame. We describe the attractor properties of the
generalized $R^p$-attractor models in the Jordan frame in terms of
the non-minimally coupled scalar theory. We also investigate the
case which leads to the standard $R^p$ inflation scenario in the
Einstein frame. We study in detail the inflationary properties of
generalized $R^p$-attractor models and we show that the
predictions are very well fitted within the Planck constraints. In
section III we discuss the problem of $R^p$ inflation in the
Jordan and Einstein frames and we pinpoint the probable reason
which leads to the conflict between the two frames. We also
discuss the difficulties of studying the generalized
$R^p$-attractor model in the Jordan frame.

Before we proceed to the core of this study, let us mention that
we shall assume that the spacetime is described by a flat
Friedmann-Robertson-Walker metric, with the line element being,
\be \label{metricfrw} ds^2 = - dt^2 + a(t)^2 \sum_{i=1,2,3}
\left(dx^i\right)^2\, , \ee where $a(t)$ denotes as usual the
scale factor. Furthermore, we assume that the metric connection is
a metric compatible, symmetric, and torsion-less affine
connection, the Levi-Civita connection. For the FRW metric, the
Ricci scalar becomes,
\begin{equation}
\label{ricciscal} R=6(2H^2+\dot{H})\, ,
\end{equation}
where $H$ denotes the Hubble rate $H=\dot{a}/a$. Also we use the
natural units physical system.

\section{Generalized $R^p$-attractors and $F(R)$ Gravity Description}

In this section we shall discuss the essential features of the
$R^p$ inflation in the Einstein frame and we shall introduce a
generalized $\alpha$-attractor-like potential which we shall call
for simplicity generalized $R^p$-attractor potential. This
Einstein frame theory corresponds to a large class of scalar
theories in the Jordan frame, and this justifies the terminology
attractor for the $\alpha$-attractor like potential which we
called generalized $R^p$-attractor potential

\subsection{$R^p$ Inflation in the Einstein Frame}

Before we get into the core of our study, let us recall the
essential features of a Jordan frame vacuum $F(R)$ gravity, for
details see the reviews \cite{reviews1,reviews2,reviews3}. The
Jordan frame action for vacuum $F(R)$ gravity is,
\begin{equation}\label{pure}
\mathcal{S}=\frac{1}{2}\int\mathrm{d}^4x \sqrt{-\hat{g}}F(R)\, ,
\end{equation}
with $\hat{g}_{\mu \nu}$ being the Jordan frame metric tensor.
Upon introducing the auxiliary scalar field $A$ for the Jordan
frame action (\ref{pure}), the gravitational action takes the
following form,
\begin{equation}\label{action1dse111}
\mathcal{S}=\frac{1}{2}\int \mathrm{d}^4x\sqrt{-\hat{g}}\left (
F'(A)(R-A)+F(A) \right )\, .
\end{equation}
Upon variation of the action (\ref{action1dse111}) with respect to
the auxiliary scalar field $A$, one obtains the solution $A=R$,
which proves the mathematical equivalence of the gravitational
actions (\ref{action1dse111}) and (\ref{pure}).

In order to obtain the Einstein frame scalar field potential
corresponding to the $F(R)$ gravity, we perform a conformal
transformation,
\begin{equation}\label{conftransmetr}
g_{\mu \nu}=e^{-\varphi }\hat{g}_{\mu \nu }\, ,
\end{equation}
with $g_{\mu \nu}$ being the Einstein frame metric. The Jordan and
Einstein frames are connected via the canonical transformation,
\begin{equation}\label{can}
\varphi =\sqrt{\frac{3}{2}}\ln (F'(A))\, ,
\end{equation}
where $\varphi$  denotes the Einstein frame canonical scalar
field. The conformally transformed action is,
\begin{align}\label{einsteinframeaction}
& \mathcal{\tilde{S}}=\int \mathrm{d}^4x\sqrt{-g}\left (
R-\frac{1}{2}\left (\frac{F''(A)}{F'(A)}\right )^2g^{\mu \nu
}\partial_{\mu }A\partial_{\nu }A -\left (
\frac{A}{F'(A)}-\frac{F(A)}{F'(A)^2}\right ) \right ) \\ \notag &
= \int \mathrm{d}^4x\sqrt{-g}\left ( R-\frac{1}{2}g^{\mu \nu
}\partial_{\mu }\varphi\partial_{\nu }\varphi -V(\varphi )\right
)\, .
\end{align}
The corresponding canonical scalar field potential  $V(\varphi )$
in the Einstein frame is,
\begin{align}\label{potentialvsigma}
V(\varphi
)=\frac{1}{2}\left(\frac{A}{F'(A)}-\frac{F(A)}{F'(A)^2}\right)=\frac{1}{2}\left
( e^{-\sqrt{2/3}\varphi }R\left (e^{\sqrt{2/3}\varphi} \right )-
e^{-2\sqrt{2/3}\varphi }F\left [ R\left (e^{\sqrt{2/3}\varphi}
\right ) \right ]\right )\, .
\end{align}
Also the Ricci scalar in terms of the canonical scalar field can
easily be obtained by solving Eq. (\ref{can}) with respect to the
auxiliary scalar $A$, and also bearing in mind the mathematical
equivalence of $A$ and $R$. Also by using Eqs.
(\ref{potentialvsigma}) and (\ref{can}), one may obtain the actual
$F(R)$ gravity which can realize a specific scalar potential. This
can be done by differentiating both sides of Eq.
(\ref{potentialvsigma}) with respect to the Ricci scalar and also
by using
$\frac{\mathrm{d}\varphi}{\mathrm{d}R}=\sqrt{\frac{3}{2}}\frac{F''(R)}{F'(R)}$.
The resulting equation is the following,
\begin{equation}\label{solvequation}
RF_R=2\sqrt{\frac{3}{2}}\frac{\mathrm{d}}{\mathrm{d}\varphi}\left(\frac{V(\varphi)}{e^{-2\left(\sqrt{2/3}\right)\varphi}}\right)
\end{equation}
where $F_R$ stands for $F_R=\frac{\mathrm{d}F(R)}{\mathrm{d}R}$.
The above differential equation in conjunction with Eq.
(\ref{can}) basically yields the vacuum $F(R)$ gravity in the
Jordan frame which generates the canonical scalar field potential
$V(\varphi)$ in the Einstein frame. In the same vain, Eq.
(\ref{potentialvsigma}) can yield the scalar potential that
corresponds to a specific $F(R)$ gravity. Now let us use the above
relations and let us discuss the $R^p$ inflation theory in the
Einstein frame.

\subsection{Generalized $R^p$ Attractor Potential and the Attractor Property in the Scalar-Jordan Frame Theory}

For attractor models, the terminology ``attractors'' indicate that
these models belong to a class of models which produce the same
inflationary indices, see for example
\cite{alpha1,alpha2,alpha3,alpha4,alpha5,alpha6,alpha7,alpha8,alpha9}.
Many well known inflationary models belong to some attractor
potential category, like the $R^2$ \cite{starob1,starob2} and the
Higgs models \cite{higgs}. All the attractor models are basically
Einstein frame models in vacuum described of course by a minimally
coupled scalar field. In addition, these models have a Jordan
frame counterpart theory described by a non-minimally coupled
scalar field theory, but all the models also have an $F(R)$
gravity description. So in order not to confuse the Jordan frame
$F(R)$ gravity description and the Jordan frame non-minimally
coupled scalar theories, we shall refer to the latter as
``$\phi$-Jordan frame'' and to the Jordan frame $F(R)$ gravity as
``Jordan frame $F(R)$ gravity theory''.

Here we shall consider generalizations of the following Einstein
frame scalar potential,
\begin{equation}\label{rpinfini}
V(\varphi)= V_0e^{-2\sqrt{\frac{2}{3}}\kappa
\varphi}\left(e^{\sqrt{\frac{2}{3}}\kappa \varphi}
-1\right)^{\frac{p}{p-1}}\, ,
\end{equation}
which is well studied in the recent literature
\cite{Motohashi:2014tra,Renzi:2019ewp}. The scalar field $\varphi$
describes an Einstein frame minimally coupled canonical scalar
field, and also in Eq. (\ref{rpinfini}), $p$ is an arbitrary
number for the Einstein frame theory and recall $\kappa=1/M_p$
where $M_p$ is the reduced Planck mass. The Einstein frame
potential above corresponds to the following Jordan frame $F(R)$
gravity theory,
\begin{equation}\label{JordanframeFR}
F(R)=R+\beta R^p\, ,
\end{equation}
where $\beta$ is a free parameter with mass dimensions $[\beta
]=[m]^{2-2p}$. As we will show explicitly in a later section, in
order for the model (\ref{JordanframeFR}) to describe an
inflationary model in the Jordan $F(R)$ theory, the parameter $p$
has to be chosen in the range $\frac{1+\sqrt{3}}{2}<p<2$, however
in the Einstein frame such a constraint does not apply.

The generalization of the potential which we shall consider in
this work has the following form,
\begin{equation}\label{rpinfinialpha}
V(\varphi)= V_0e^{-2\sqrt{\frac{2}{3 \alpha}}\kappa
\varphi}\left(e^{\sqrt{\frac{2}{3\alpha }}\kappa \varphi}
-1\right)^{\frac{p}{p-1}}\, ,
\end{equation}
which for $\alpha=1$ becomes identical with the $R^p$ model in the
Einstein frame. Now let us demonstrate to which attractor class
does the potential (\ref{rpinfinialpha}) belong to, and to this
end, consider the following $\phi$-Jordan frame action,
\begin{equation}\label{phijordanframeaction}
\mathcal{S}_J=\int
d^4x\left(\frac{\Omega(\phi)}{2\kappa^2}R-\frac{\omega
(\phi)}{2}g^{\mu \nu} \partial_{\mu}\phi \partial_{\nu}\phi
-V_J(\phi)\right)\, ,
\end{equation}
where the scalar field describes the non-canonical Jordan frame
scalar field and $\Omega(\phi)=1+\xi f(\phi)$ with $\xi$ being an
arbitrary parameter and $f(\phi)$ and analytic function of $\phi$.
The particular choice for the function $\Omega(\phi)=1+\xi
f(\phi)$  is motivated by many non-minimally coupled theories of
gravity having this form, like the Higgs scalar tensor theory (see
Eq. (52) of \cite{Odintsov:2021nqa} and Eq. (31) of
\cite{Oikonomou:2021iid}.

 We shall call the attractor
class of the Einstein frame potential (\ref{rpinfinialpha}) as
``$R^p$-attractors'', and these are realized when the scalar
potential in the $\phi$-Jordan frame is chosen to be of the
following form,
\begin{equation}\label{potentialJordanframe}
V_J(\phi)=V_0\left(\Omega(\phi)-1 \right)^{\frac{p}{p-1}}\, ,
\end{equation}
and also the kinetic term function $\omega (\phi)$ is chosen to
be,
\begin{equation}\label{kinetictermfunction}
\omega (\phi)=\frac{1}{4\xi}\frac{\left(\frac{d \Omega (\phi)}{d
\phi}\right)^2}{\Omega(\phi)}\, .
\end{equation}
Thus basically, the $R^p$-attractors correspond the choices
(\ref{potentialJordanframe}) and (\ref{kinetictermfunction}).
Performing the following conformal transformation on the Jordan
frame metric $g_{\mu \nu}$,
\begin{equation}\label{conformaltransmetr}
\tilde{g}_{\mu \nu}=\Omega(\phi)g_{\mu \nu}\, ,
\end{equation}
where $\tilde{g}_{\mu \nu}$ is the Einstein frame metric tensor,
we obtain the following Einstein frame action,
\begin{equation}\label{alphaact}
\mathcal{S}_E=\sqrt{-\tilde{g}}\left(\frac{\tilde{R}}{2\kappa^2}-\tilde{g}^{\mu
\nu}\partial_{\mu}\varphi
\partial_{\nu}\varphi-V(\varphi)\right)\,
,
\end{equation}
where the ``tilde'' denotes Einstein frame quantities, and the
Einstein frame potential $V(\phi)$ is related with the Jordan
frame potential $V_J(\phi)$ in the following way,
\begin{equation}\label{jordaneinsteinframepotential}
V(\varphi)=\Omega^{-2}(\phi)V_J(\phi)\, .
\end{equation}
The general relation between the Jordan frame scalar field $\phi$
and the canonical Einstein frame scalar field $\varphi$ is
\cite{newsergei},
\begin{equation}\label{generalrelationbetweenscalar}
\left( \frac{d \varphi}{d
\phi}\right)^{2}=\frac{3}{2}\frac{\left(\frac{d \Omega (\phi)}{d
\phi}\right)^2}{\Omega(\phi)}+\frac{\omega (\phi)}{\Omega(\phi)}\,
,
\end{equation}
so for the case of the generalized $R^p$-attractors for which the
kinetic term function $\omega (\phi)$ is given by Eq.
(\ref{kinetictermfunction}), by integrating the relation between
the Jordan frame scalar field $\phi$ and the canonical Einstein
frame scalar field $\varphi$, namely Eq.
(\ref{generalrelationbetweenscalar}), we get,
\begin{equation}\label{omegaphigeneral}
\Omega (\phi)=e^{\sqrt{\frac{2}{3\alpha}}\varphi}\, ,
\end{equation}
where the parameter $\alpha$ is defined to be,
\begin{equation}\label{alphageneraldefinition}
\alpha=1+\frac{1}{6\xi}\, .
\end{equation}
It is apparent that by substituting Eq. (\ref{omegaphigeneral}) in
Eq. (\ref{jordaneinsteinframepotential}) one obtains the
generalized $R^p$-attractor potential of Eq.
(\ref{rpinfinialpha}).

Before we proceed, let us discuss several important issues related
to the generalized $R^p$-attractor. Firstly, in the Einstein frame
there is no constraint on the parameter $p$, in contrast to the
Jordan frame $F(R)$ description of $R^p$ inflation, in which case
in order for having an inflationary description, the parameter $p$
is constrained to be $\frac{1+\sqrt{3}}{2}<p<2$. Secondly, the
$R^p$ theory is not a viable inflationary theory in the $F(R)$
Jordan frame, in contrast to its Einstein frame counterpart theory
with potential (\ref{rpinfini}), as it can be seen in Refs.
\cite{Motohashi:2014tra,Renzi:2019ewp}. Furthermore, we abusively
used the terminology $R^p$-attractors to characterize the class of
attractor potentials (\ref{rpinfinialpha}), since the Einstein
frame potential (\ref{rpinfinialpha}) does not originate from an
$R^p$ theory in the Jordan $F(R)$ frame. In fact it is impossible
to find the $F(R)$ Jordan frame theory for a general $p$ and can
be found for specific values of $\alpha$ and $p$, without these
values guaranteeing the viability of the inflationary theory (see
the Appendix A for an explicit example for which the $F(R)$
gravity can be evaluated). Moreover, the case $\alpha=1$ can be
realized when $\xi\to \infty$, or equivalently when $\Omega
(\phi)\ll \frac{3}{2} \frac{\left(\frac{d \Omega (\phi)}{d
\phi}\right)^2}{\omega(\phi)} $ (see for example Ref.
\cite{alpha12}). The $\phi$-Jordan frame counterpart theory to the
potential (\ref{rpinfinialpha}) will yield the same inflationary
phenomenology with the Einstein frame theory with the potential
(\ref{rpinfinialpha}) for a general $\alpha$. However for
$\alpha=1$, the same does not apply to the $F(R)$-Jordan frame
theory, which as we show in a later section is not a viable
inflationary theory for any value of the parameter $p$. This is a
peculiar result, since by the conformal equivalence the $\alpha=1$
theory with potential (\ref{rpinfini}) and the $R^p$ theory should
be equivalent. This however is not true, and in a later section we
shall show that this is due to the peculiarity of the actual $R^p$
inflation scenario in the Jordan frame.

Let us proceed to the phenomenology of the generalized
$R^p$-attractor potential (\ref{rpinfinialpha}) in some detail.
The results should coincide with the phenomenology of the model
(\ref{rpinfini}) studied in Refs.
\cite{Motohashi:2014tra,Renzi:2019ewp} for $\alpha=1$, which in
the case of the generalized $R^p$-attractors is obtained in the
limit $\xi \to \infty$. To start with, let us calculate the
spectral indices for the generalized $R^p$-attractor potential,
and since the theory is a minimally coupled scalar field theory,
the slow-roll indices are given by (see for example the review
\cite{reviews1}),
\begin{equation}\label{epsilonslow}
\epsilon=\frac{1}{2\kappa^2}\left(\frac{V'}{V} \right)^2\, ,
\end{equation}
\begin{equation}\label{etaslow}
\eta=\frac{1}{\kappa^2}\frac{V''}{V}\, ,
\end{equation}
while the spectral index of the primordial scalar perturbations
has the form,
\begin{equation}\label{spectralindexscalar}
n_s=1-6\epsilon+2\eta\, ,
\end{equation}
and the tensor-to-scalar ratio is,
\begin{equation}\label{tensortoscalarratio}
r=16\epsilon\, .
\end{equation}
The $e$-foldings number can be expressed in terms of the scalar
field potential as follows,
\begin{equation}\label{efoldingsnumber}
N=\kappa^2\int^{\phi_i}_{\phi_f}\frac{V}{V'}d\phi\, ,
\end{equation}
where $\phi_i$ is the value of the scalar field when the
corresponding scalar perturbation mode exits the horizon at the
beginning of the inflationary era and $\phi_f$ is the value of the
scalar field at the end of the inflationary era. Let us proceed to
calculate the inflationary parameters and observational indices in
detail for the generalized $R^p$-attractor potential
(\ref{rpinfinialpha}). Firstly let us quote the expressions for
the slow-roll indices and these are,
\begin{equation}\label{slowroll1exp}
\epsilon=\frac{\left((p-2) e^{\sqrt{\frac{2}{3}}
\sqrt{\frac{1}{\alpha }} \kappa  \varphi }-2 p+2\right)^2}{3
\alpha  (p-1)^2 \left(e^{\sqrt{\frac{2}{3}} \sqrt{\frac{1}{\alpha
}} \kappa  \varphi }-1\right)^2}\, ,
\end{equation}
\begin{equation}\label{slowroll2}
\eta=\frac{2 \left(\left(-5 p^2+13 p-8\right)
e^{\sqrt{\frac{2}{3}} \sqrt{\frac{1}{\alpha }} \kappa  \varphi
}+(p-2)^2 e^{2 \sqrt{\frac{2}{3}} \sqrt{\frac{1}{\alpha }} \kappa
\varphi }+4 (p-1)^2\right)}{3 \alpha  (p-1)^2
\left(e^{\sqrt{\frac{2}{3}} \sqrt{\frac{1}{\alpha }} \kappa
\varphi }-1\right)^2}\, .
\end{equation}
Having these available the final value of the scalar field at the
end of inflation $\varphi_f$ is found by solving
$\epsilon(\varphi)=\mathcal{O}(1)$, and we get,
\begin{equation}\label{varphiend}
\varphi_f=\frac{\sqrt{\frac{3}{2}} \ln \left(\frac{3 \alpha
+(3\alpha -2) p^2+\sqrt{3} \sqrt{\alpha  (p-1)^2 p^2}-6 (\alpha
-1) p-4}{3 \alpha +(3 \alpha -1) p^2+(4-6 \alpha )
p-4}\right)}{\sqrt{\frac{1}{\alpha }} \kappa }\, ,
\end{equation}
and by using Eq. (\ref{efoldingsnumber}) we can easily find the
value of the scalar field at the first horizon crossing at the
beginning of inflation $\varphi_i$ which is,
\begin{equation}\label{varphi}
\varphi_i=\frac{\sqrt{\frac{3}{2}} \ln \left(\frac{e^{-\frac{4
(p-2) N}{3 \alpha  p}} \left((p-2) e^{\sqrt{\frac{2}{3}}
\sqrt{\frac{1}{\alpha }} \kappa  \varphi_f}-2 p+2\right)+2
p-2}{p-2}\right)}{\sqrt{\frac{1}{\alpha }} \kappa } \, .
\end{equation}
Accordingly, the spectral index as a function of the scalar field
is,
\begin{equation}\label{spectralnsexplicit}
n_s=\frac{\left(3 \alpha +(3 \alpha -2) p^2+(8-6 \alpha )
p-8\right) e^{2 \sqrt{\frac{2}{3}} \sqrt{\frac{1}{\alpha }} \kappa
\varphi }-2 (p-1) (-3 \alpha +(3 \alpha -2) p+8)
e^{\sqrt{\frac{2}{3}} \sqrt{\frac{1}{\alpha }} \kappa  \varphi
}+(3 \alpha -8) (p-1)^2}{3 \alpha  (p-1)^2
\left(e^{\sqrt{\frac{2}{3}} \sqrt{\frac{1}{\alpha }} \kappa
\varphi }-1\right)^2}\, ,
\end{equation}
and the tensor-to-scalar ratio as a function of the scalar field
is,
\begin{equation}\label{tensotoscalarasfunctionofscalarfield}
r=\frac{16 \left((p-2) e^{\sqrt{\frac{2}{3}} \sqrt{\frac{1}{\alpha
}} \kappa  \varphi }-2 p+2\right)^2}{3 \alpha  (p-1)^2
\left(e^{\sqrt{\frac{2}{3}} \sqrt{\frac{1}{\alpha }} \kappa
\varphi }-1\right)^2}\, .
\end{equation}
Hence having these analytic relations available we can now proceed
with the confrontation of the theory with the Planck 2018 data
\cite{Planck:2018jri} which constrain the spectral index and the
tensor-to-scalar ratio as follows,
\begin{equation}\label{planck2021}
n_s=0.962514\pm 0.00406408,\,\,\,r<0.064\, .
\end{equation}
Firstly, let us give several values on the free parameters
$\alpha$ and $p$ in order to see how the phenomenology behaves
depending on the values of these free parameters. For both the
indices, the values of $p$ and $\alpha$ that yield a viable
phenomenology are in the ranges $1.9<p<2.01$ and $1\leq \alpha\leq
8$.
\begin{figure}
\centering
\includegraphics[width=20pc]{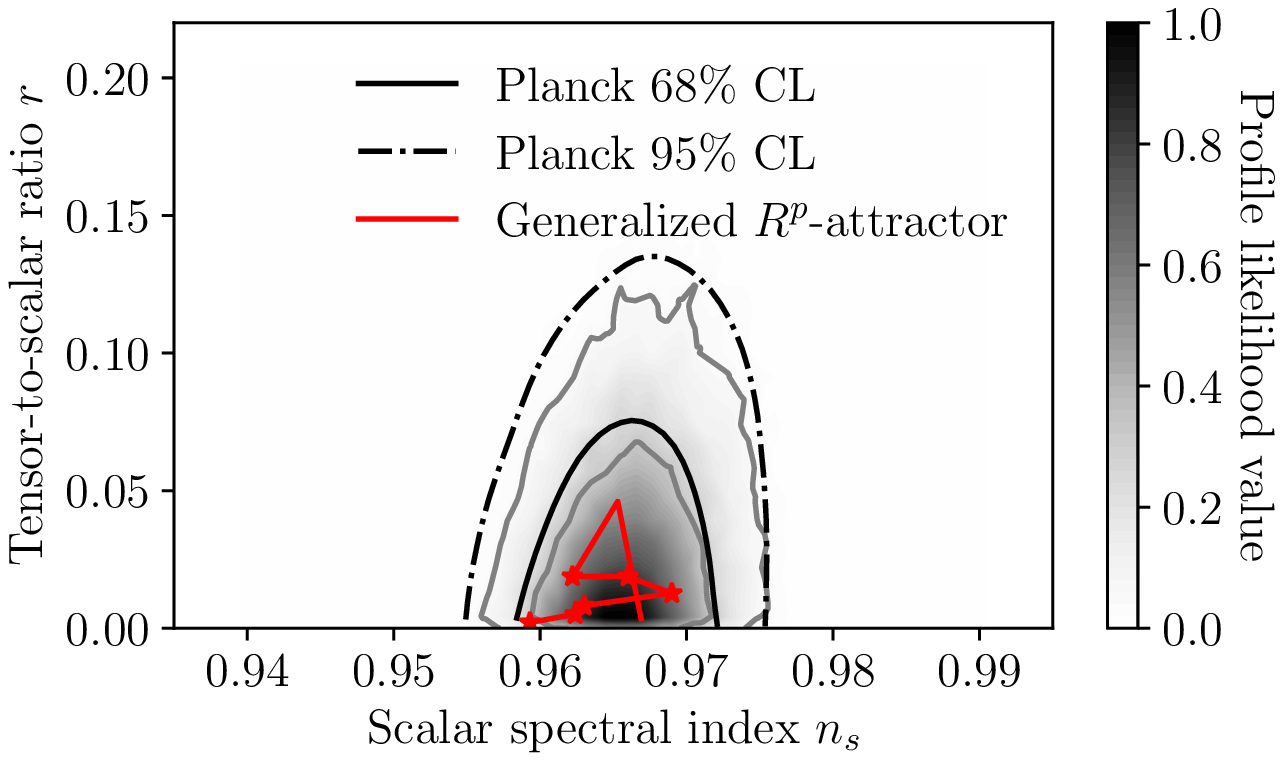}
\includegraphics[width=20pc]{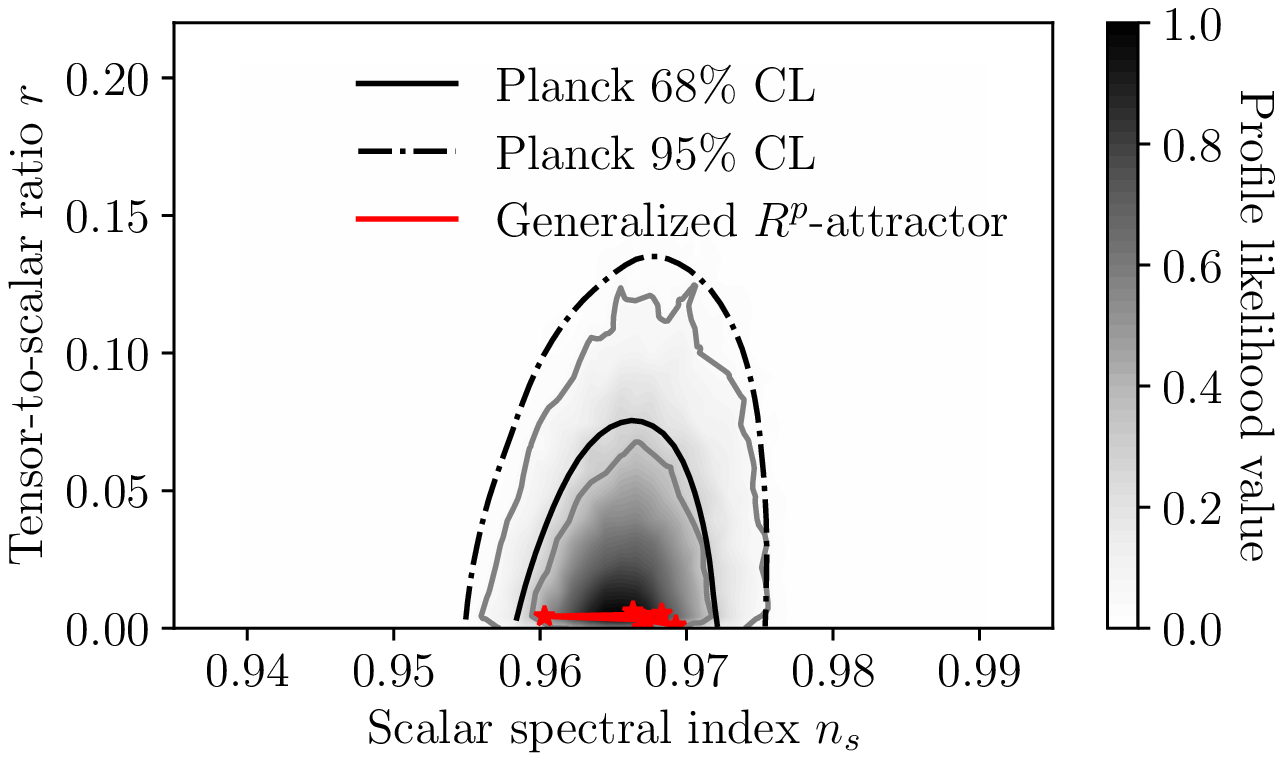}
\includegraphics[width=20pc]{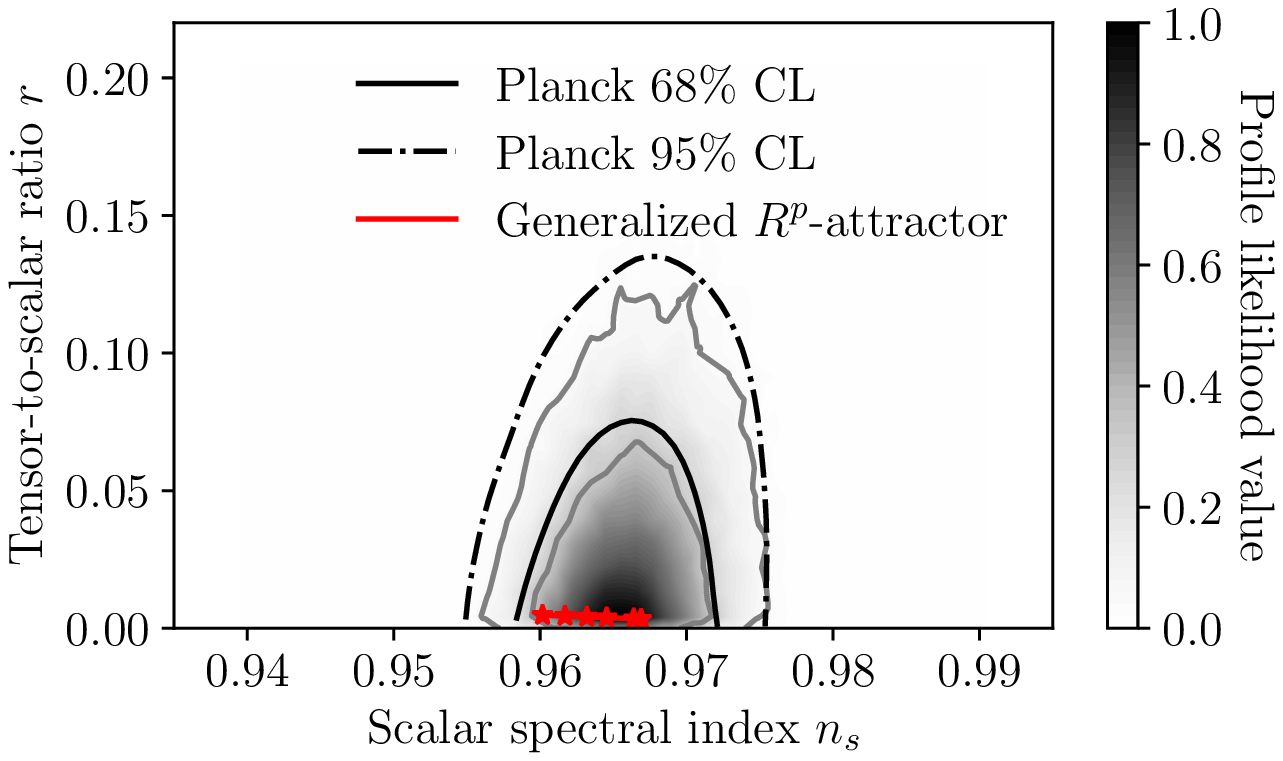}
\caption{The generalized $R^p$-attractor model of Eq. (12) versus
the Planck likelihood curves. In the upper left plot we used
$1.9\leq p \leq 2.01$ and $1\leq \alpha\leq 8$ for $N=60$, in the
upper right plot we used $1.9\leq p \leq 1.9999$ and $0.1\leq
\alpha\leq 0.9$ for $N=50$, whereas in the bottom plot we used
$1.9\leq p \leq 1.9999$ and $\alpha=1$ for
$N=[50,60]$.}\label{plot1}
\end{figure}
Also small values of the parameter $\alpha\leq 1$ significantly
affect the tensor-to-scalar ratio which render it small, however
most of these values strongly affect the spectral index making it
to have values non compatible with the Planck data. If however,
$p$ takes values quite close to $p\simeq 2$, for example
$p=1.9999$ and $0.1\leq \alpha \leq 0.9$ then the tensor-to-scalar
becomes quite small and the spectral index is compatible with the
Planck 2018 data. The results of our analysis where we confront
the model for various parameter values can be found in Fig.
\ref{plot1}, where we present the model's predictions compared to
the Planck likelihood curves. In the upper left plot of Fig.
\ref{plot1} we used $1.9\leq p \leq 2.01$ and $1\leq \alpha\leq 8$
for $N=60$, in the upper right plot we used $1.9\leq p \leq
1.9999$ and $0.1\leq \alpha\leq 0.9$ for $N=50$, whereas in the
bottom plot we used $1.9\leq p \leq 1.9999$ and $\alpha=1$ for
$N=[50,60]$. As it can be seen in all cases, the model is quite
well fitted within the Planck likelihood curves, and in some cases
the model is found at the sweet spot of the Planck likelihood
curves.

\section{$R^p$ Gravity in the Jordan Frame: A Difference on Inflationary Dynamics}

In this section we shall reveal an important feature of the $R^p$
inflation in the $f(R)$ Jordan frame. In the previous section we
demonstrated that the $R^p$ inflation is a viable theory in the
Einstein frame, for various values of the parameter $\alpha$,
including the value $\alpha=1$, and the model was found to be
compatible with the Planck data. The same applies for the
$\phi$-Jordan frame theory described by the non-minimally coupled
generalized $R^p$-attractors. However, as we now demonstrate
shortly, the $R^p$ inflation in the $f(R)$ Jordan frame is not a
viable inflationary theory, at least when studied in the formal
way appearing in the literature. The focus in this section is
investigating the reason why this peculiarity occurs, since the
scalar spectral index and the tensor-to-scalar ratio are conformal
invariant quantities, thus one would expect that these should be
the same in all conformally related frames. As we will see, the
incompatibility between the two frames is possibly due to a
simplification made in the $f(R)$ Jordan frame for the sake of
analyticity. A similar discussion for the complexity of the Jordan
frame counterpart theories, can be found in the Appendix B where a
detailed study of the $\phi$-Jordan frame theory is presented in
detail. Consider the $R^p$ inflationary theory in the $f(R)$
Jordan frame,
\begin{equation}\label{polynomialfr}
f(R)=R+\beta R^p\, ,
\end{equation}
with $p\neq 2$. The Friedman equation of  $f(R)$ gravity is,
\begin{equation}\label{friedmannewappendix}
3 H^2F=\frac{RF-f}{2}-3H\dot{F}\, ,
\end{equation}
with $F=\frac{\partial f}{\partial R}$. During the inflationary
era, we have approximately $F\sim p\beta R^{p-1}$, at leading
order, so the Friedman equation (\ref{friedmannewappendix}) reads,
\begin{align}\label{eqnsofmkotionfrpolyappendix}
& 3 H^2p\beta R^{p-1}=\frac{\beta (p-1)R^{p-1}}{2}-3p(p-1)\beta
HR^{p-2}\dot{R}\, .
\end{align}
Also $R=12H^2+6\dot{H}$, which during the slow-roll inflationary
era becomes at leading order $R\sim 12 H^2$ and $\dot{R}\sim
24H\dot{H}$, since the slow-roll assumptions in the $f(R)$ Jordan
frame are,
\begin{equation}\label{slowrolljordan}
\dot{H}\ll H^2,\,\,\,\ddot{H}\ll H\dot{H}\, ,
\end{equation}
and therefore the Friedman equation takes the following form at
leading order,
\begin{equation}\label{leadingordereqnappendix}
3H^2p\beta \simeq 6\beta (p-1)H^2-6p\beta(p-1)\dot{H}+3\beta
(p-1)\dot{H}\, ,
\end{equation}
which by solving it, yields the following solution for the Hubble
rate,
\begin{equation}\label{hubblefrpolyappendix}
H(t)=\frac{-2p^2+3p-1}{(p-2)t}\, .
\end{equation}
The evolution in Eq. (\ref{hubblefrpolyappendix}) can describe an
inflationary evolution in the $f(R)$ Jordan frame only if the
parameter $p$ takes values in the range
$\frac{1+\sqrt{3}}{2}<p<2$. This is in contrast to the Einstein
frame theory in which case the parameter $p$ is unconstrained.
Using the evolution of Eq. (\ref{hubblefrpolyappendix}), we can
obtain the slow-roll indices for the power law $f(R)$ gravity
model (\ref{polynomialfr}), and these have the following form,
\begin{equation}\label{epsiloniforfrpoly}
\epsilon_1=\frac{p-2}{1-3p+2p^2},\,\,\,\epsilon_2\simeq
0,\,\,\,\epsilon_3=(p-1)\epsilon_1,\,\,\,\epsilon_4=\frac{p-2}{p-1}\,
.
\end{equation}
The spectral index of scalar perturbations and the
tensor-to-scalar ratio for a vacuum $f(R)$ gravity in the Jordan
frame have the following form \cite{reviews1,reviews2},
\begin{equation}\label{vacuumspectral}
n_s=1-6\epsilon_1-2\epsilon_4,\,\,\,r=48 \epsilon_1^2\, .
\end{equation}
By performing a simple analysis, it is easy to show that the only
value of the parameter $p$ that renders the scalar spectral index
compatible with the Planck data is $p=1.81$, however the model is
not viable since for that value of $p$ the tensor-to-scalar ratio
takes the value $r=0.13$. This is in contrast to the Einstein
frame theory in which case the model was found viable in the
previous section, for $\alpha=1$. We think that the reason for
this inconsistency is the following: it has to do with the
approximation we made $F\sim p\beta R^{p-1}$, which we applied it
in order to obtain analytic results.

Now let us use another formalism in order to study the $R^p$
inflation in the $f(R)$ Jordan frame, this will also provide
useful insights regarding the inconsistency in the two frames. As
we now show, using the formalism we shall now present, the $R^p$
inflation in the $f(R)$ Jordan frame is a viable theory compatible
with the Planck data. This result seems to be correct and
perfectly aligned with the principle which dictates that
conformally related frames should yield identical results for
conformal invariant quantities. Let us quote for convenience again
the field equations, which for the FRW metric read,
\begin{align}
\label{JGRG15} 0 =& -\frac{f(R)}{2} + 3\left(H^2 + \dot H\right)
F_R(R) - 18 \left( 4H^2 \dot H + H \ddot H\right) F_{RR}(R)\, ,\\
\label{Cr4b} 0 =& \frac{f(R)}{2} - \left(\dot H +
3H^2\right)F_R(R) + 6 \left( 8H^2 \dot H + 4 {\dot H}^2 + 6 H
\ddot H + \dddot H\right) F_{RR}(R) + 36\left( 4H\dot H + \ddot
H\right)^2 F_{RRR} \, ,
\end{align}
where $F_{RR}=\frac{\mathrm{d}^2f}{\mathrm{d}R^2}$, and
$F_{RRR}=\frac{\mathrm{d}^3f}{\mathrm{d}R^3}$. Also the Ricci
scalar as a function of the Hubble rate for the FRW metric is
$R=12H^2 + 6\dot H$. We shall now derive the $n_s-r$ relation for
$f(R)$ gravity and we shall apply the formalism for the power-law
$f(R)$ gravity of Eq. (\ref{polynomialfr}). Let us recall the
general form of the slow-roll indices relevant for studying $f(R)$
gravity inflation in vacuum, namely $\epsilon_1$ ,$\epsilon_2$,
$\epsilon_3$, $\epsilon_4$, which are
\cite{Hwang:2005hb,reviews1},
\begin{equation}
\label{restofparametersfr}\epsilon_1=-\frac{\dot{H}}{H^2}, \quad
\epsilon_2=0\, ,\quad \epsilon_3= \frac{\dot{F}_R}{2HF_R}\, ,\quad
\epsilon_4=\frac{\ddot{F}_R}{H\dot{F}_R}\,
 .
\end{equation}
Now, let us assume that $\epsilon_i\ll 1$, $i=1,3,4$, and the
scalar spectral index and the tensor-to-scalar ratio are
\cite{reviews1,Hwang:2005hb,Odintsov:2020thl},
\begin{equation}
\label{epsilonall} n_s=
1-\frac{4\epsilon_1-2\epsilon_3+2\epsilon_4}{1-\epsilon_1},\quad
r=48\frac{\epsilon_3^2}{(1+\epsilon_3)^2}\, .
\end{equation}
Now, the Raychaudhuri equation dictates that,
\begin{equation}\label{approx1}
\epsilon_1=-\epsilon_3(1-\epsilon_4)\, ,
\end{equation}
and due to the slow-roll assumption for the slow-roll indices, we
have,
\begin{equation}
\label{spectralfinal} n_s\simeq 1-6\epsilon_1-2\epsilon_4\, .
\end{equation}
Furthermore regarding the tensor-to-scalar ratio, we have $r\simeq
48 \epsilon_3^2$, and due to the fact that $\epsilon_1\simeq
-\epsilon_3$, we finally have,
\begin{equation}
\label{tensorfinal} r\simeq 48\epsilon_1^2\, .
\end{equation}
The analytic calculation of the slow-roll index
$\epsilon_4=\frac{\ddot{F}_R}{H\dot{F}_R}$ is of profound
importance in order to obtain the $n_s-r$ relation, so let us
express it in terms of the slow-roll index $\epsilon_1$, and we
find,
\begin{equation}\label{epsilon41}
\epsilon_4=\frac{\ddot{F}_R}{H\dot{F}_R}=\frac{\frac{d}{d
t}\left(F_{RR}\dot{R}\right)}{HF_{RR}\dot{R}}=\frac{F_{RRR}\dot{R}^2+F_{RR}\frac{d
(\dot{R})}{d t}}{HF_{RR}\dot{R}}\, .
\end{equation}
We need to find  $\dot{R}$, so by using the slow-roll assumption
$\ddot{H}\ll H \dot{H}$, we have,
\begin{equation}\label{rdot}
\dot{R}=24\dot{H}H+6\ddot{H}\simeq 24H\dot{H}=-24H^3\epsilon_1\, .
\end{equation}
Now upon combining Eqs. (\ref{rdot}) and (\ref{epsilon41}), after
some algebra we obtain,
\begin{equation}\label{epsilon4final}
\epsilon_4\simeq -\frac{24
F_{RRR}H^2}{F_{RR}}\epsilon_1-3\epsilon_1+\frac{\dot{\epsilon}_1}{H\epsilon_1}\,
,
\end{equation}
and due to the fact that the term $\dot{\epsilon}_1$ is,
\begin{equation}\label{epsilon1newfiles}
\dot{\epsilon}_1=-\frac{\ddot{H}H^2-2\dot{H}^2H}{H^4}=-\frac{\ddot{H}}{H^2}+\frac{2\dot{H}^2}{H^3}\simeq
2H \epsilon_1^2\, ,
\end{equation}
the slow-roll index $\epsilon_4$ reads,
\begin{equation}\label{finalapproxepsilon4}
\epsilon_4\simeq -\frac{24
F_{RRR}H^2}{F_{RR}}\epsilon_1-\epsilon_1\, .
\end{equation}
Now we shall introduce the following dimensionless parameter $x$,
\begin{equation}\label{parameterx}
x=\frac{48 F_{RRR}H^2}{F_{RR}}\, ,
\end{equation}
so the slow-roll index $\epsilon_4$ can be written in terms of $x$
as follows,
\begin{equation}\label{epsilon4finalnew}
\epsilon_4\simeq -\frac{x}{2}\epsilon_1-\epsilon_1\, .
\end{equation}
Now, by combining Eqs. (\ref{epsilon4finalnew}) and
(\ref{spectralfinal}), we can directly express the spectral index
as a function of the first slow-roll index $\epsilon_1$ as
follows,
\begin{equation}\label{asxeto1}
n_s-1=-4\epsilon_1+x\epsilon_1\, .
\end{equation}
By eliminating $\epsilon_1$ and combining Eqs. (\ref{asxeto1}) and
(\ref{tensorfinal}), we finally obtain,
\begin{equation}\label{mainequation}
r\simeq \frac{48 (1-n_s)^2}{(4-x)^2}\, ,
\end{equation}
which is a quite valuable relation for vacuum $f(R)$ gravity, and
it is derived solely on the slow-roll assumptions. Now, let us
consider the $f(R)$ gravity of relation (\ref{polynomialfr}) and
let us calculate the $n_s-r$ relation for this $f(R)$ gravity. The
crucial parameter to be calculated is the dimensionless parameter
$x$, which for the particular $f(R)$ gravity under study reads,
\begin{equation}\label{xfrpower}
x=\frac{48(p-2) H^2}{R}\, .
\end{equation}
Recall that $R=12H^2+6\dot{H}$ for a flat FRW metric, so due to
the slow-roll assumption, $R\sim 12 H^2$, hence the parameter $x$
is constant for the $f(R)$ gravity at hand and it is equal to
$x=4(p-2)$. Thus, the $n_s-r$ relation of Eq. (\ref{mainequation})
becomes,
\begin{equation}\label{nsrfinal}
r=3\frac{\left(1-n_s \right)^2}{\left(3-p \right)^2}\, .
\end{equation}
The above relation can generate a viable phenomenology for several
values of $p$ if $n_s$ is within the acceptable parameter values
ranges of the Planck data. For example, if $n_s=0.961$ for
$p=1.81$ we get $r=0.00322223$, which is well fitted with the
Planck observational data. This result is in more agreement with
the results of the previous section, since the $R^p$ theory in the
$f(R)$ Jordan frame is viable. In general, according to our
approach, if the spectral index is compatible with the
observational data, then the tensor-to-scalar ratio is also
compatible with the observations, a feature absent in the other
approach presented earlier. However, this result should be
discussed in the context of both formalisms presented in this
section, in order to pinpoint the source of the inconsistency. If
for example, the first formalism is assumed to hold true, then the
second formalism cannot be used since $\epsilon_1=$const. However,
we believe that the second formalism is the correct approach and
thus the $R^p$ theory in the $f(R)$ Jordan frame is viable, a
result which is also supported by the viability of the Einstein
frame counterpart theory. The problem with the first formalism
which we presented in the first part of this section, is mainly
the assumption $F_R\sim \beta n R^{p-1}$ which we did for the sake
of analyticity and only for that. It seems therefore that the
$R^p$ theory in the $f(R)$ Jordan frame is a viable inflationary
theory, contrary to what was believed to date. This result is
further supported by the fact that the counterpart theory in the
Einstein frame is viable, as it should be due to the conformal
relation between the two frames. It can be checked that the
Einstein frame counterpart scalar theory yields almost identical
results with the Jordan frame theory (\ref{nsrfinal}). Also
another fact that further supports the second formalism we
presented for the $R^p$ theory in the $f(R)$ Jordan frame is the
irrelevance of the values of the parameter $p$ in order to obtain
a viable inflationary theory, in contrast to the first formalism,
in which case an inflationary theory occurs for
$\frac{1+\sqrt{3}}{2}<p<2$. In the Einstein frame theory inflation
depends on the parameter $p$ but the values of $p$ are not
directly constrained.

As a final comment, the differences in the observational indices
in conformally related theories, are mainly due to the different
approximations used for the sake of lack of analyticity. In the
Appendix B we shall present the $\phi$-Jordan frame inflationary
scalar theory and we shall show that it is quite hard to be
studied analytically, without making several assumptions. These
assumptions might affect the final forms of the observational
indices.

\section{Conclusions}

In this paper we studied a new class of cosmological inflationary
attractors, which we named generalized $R^p$-attractors, since the
Einstein frame scalar potential is a generalization of the
Einstein frame scalar potential corresponding to the Jordan frame
$R^p$ gravity. We presented several features of the corresponding
non-minimally coupled scalar field theory that yield an attractor
type property, since a large class of models with an arbitrary
non-minimal coupling having specific properties, yields the same
Einstein frame scalar potential and consequently the same
inflationary phenomenology. We studied the inflationary
phenomenology of the resulting Einstein frame scalar field theory
and as we showed, the viability of the generalized
$R^p$-attractors is guaranteed for a large range of the two
parameters that characterize the model. Also we discussed an
important issue related to the $R^p$ inflation in the Jordan and
Einstein frame. The $R^p$ inflation in the Einstein frame is a
viable theory and this is in contrast to the well-known
non-viability of the standard power-law $R^p$ model in the Jordan
frame. Following the general principle that conformal invariant
quantities between conformally related frames, should be identical
when calculated in the two frames, the result of $R^p$ inflation
between the two frames is rather unappealing. We pinpoint the
reason why $R^p$ inflation in the Jordan frame is rendered
non-viable, and it seems that it has to do with an approximation
made for the sake of analyticity. We presented a formalism that
may alleviate this problem, and as we showed, by using the
slow-roll conditions solely, the Jordan frame $R^p$ theory can be
viable. To be more accurate, if the spectral index of the $R^p$
theory is compatible with the Planck data, then the
tensor-to-scalar ratio is also viable, based on our approach. The
whole problem arises to our opinion, due to the lack of
analyticity and the complexity of the Jordan frame theory. We also
present the inflationary phenomenology of the non-minimally
coupled theory in the Jordan frame, which further shows how
difficult it is to obtain analytic results in the Jordan frame.
However, by conformal invariance of the spectral index and of the
tensor-to-scalar ratio, the Einstein frame study suffices for the
inflationary study. Finally, the same consideration made in this
article can be extended to Einstein-Gauss-Bonnet inflationary
theories, a task for which work is in progress.

\section*{Acknowledgments}

This work was supported by MINECO (Spain), project
PID2019-104397GB-I00 (S.D.O). This work by S.D.O was also
partially supported by the program Unidad de Excelencia Maria de
Maeztu CEX2020-001058-M, Spain.

\section*{Appendix A: An Explicit Example for which the $F(R)$ Gravity Can be Obtained}

In this Appendix we shall evaluate the $f(R)$ gravity which can
generate the generalized $R^p$-attractor potential
(\ref{rpinfinialpha}) for specific values of the parameters
$\alpha$ and $p$. Apart from the value $\alpha=1$, one case for
which we can extract analytical results is the case $\alpha=4$ and
$p=3/2$ in which case the equation (\ref{solvequation}) takes the
form,
\begin{equation}\label{exactformanalyticalgebraic}
f_R R-4 \left(\sqrt{f_R}-1\right)^2 \left(\frac{5
\sqrt{f_R}}{4}-\frac{1}{2}\right) f_R V_0=0\, ,
\end{equation}
where recall that $f_R=\frac{d f}{d R}$, which when solved it
yields,
\begin{align}\label{frderivativeanalytic}
& f_R=\frac{18}{25}-\frac{\sqrt[3]{2} \left(-1800 R V_0^3-441
V_0^4\right)}{225 V_0^2 \sqrt[3]{625 R^2 V_0^4+25 \sqrt{625 R^4
V_0^8-24600 R^3 V_0^9+244020 R^2 V_0^{10}-38416 R V_0^{11}}+13300
R V_0^5-686 V_0^6}}\\ \notag & +\frac{\sqrt[3]{625 R^2 V_0^4+25
\sqrt{625 R^4 V_0^8-24600 R^3 V_0^9+244020 R^2 V_0^{10}-38416 R
V_0^{11}}+13300 R V_0^5-686 V_0^6}}{25 \sqrt[3]{2} V_0^2}\, ,
\end{align}
and this is the only thing that can analytically be evaluated.
Thus in conclusion, the $f(R)$ Jordan frame evaluation and study
of inflation is particularly difficult to be performed
analytically, so one must rely on the Einstein frame study and the
conformal equivalence of the two frames in order to study the
inflationary dynamics of generalized $R^p$ inflation.

\section*{Appendix B: Inflationary Dynamics in the $\phi$-Jordan Frame}

In this section we shall present the field equations and the
formalism of inflationary dynamics in the $\phi$-Jordan frame for
the generalized $R^p$ inflation. As we shall demonstrate, as in
the case presented in the Appendix A, the study of generalized
$R^p$ inflation in the $\phi$-Jordan frame is particularly
difficult to study analytically, thus one must rely on the
Einstein frame calculation and the conformal equivalence of the
frames in order to study generalized $R^p$ inflation analytically.
To start with, consider the gravitational action of a the
non-minimally coupled scalar field in the $\phi$-Jordan frame,
\begin{equation}\label{phijordanframeactionappendixB}
\mathcal{S}_J=\int
d^4x\left(\frac{\Omega(\phi)}{2\kappa^2}R-\frac{\omega
(\phi)}{2}g^{\mu \nu} \partial_{\mu}\phi \partial_{\nu}\phi
-V_J(\phi)\right)\, ,
\end{equation}
and the slow-roll indices are defined as follows,
\cite{Hwang:2005hb,reviews1},
\begin{equation}
\label{restofparametersfrappendixB}\epsilon_1=-\frac{\dot{H}}{H^2},
\quad \epsilon_2=\frac{\ddot{\phi}}{H\dot{\phi}}\, ,\quad
\epsilon_3= \frac{\dot{\Omega}(\phi)}{2H\Omega (\phi)}\, ,\quad
\epsilon_4=\frac{\ddot{E}(\phi)}{H\dot{E}(\phi)}\,
 ,
\end{equation}
where the function $E(\phi)$ is defined as follows,
\begin{equation}\label{epsilonappendix}
E(\phi)=\frac{\Omega (\phi)}{\kappa^2\dot{\phi}^2}\left( \omega
(\phi)\dot{\phi}^2+\frac{3\dot{\Omega}(\phi)^2}{2\Omega
(\phi)\kappa^2} \right)\, .
\end{equation}
In the present case, the field equations read,
\begin{equation}\label{eq1appendix}
3\Omega (\phi)=\frac{1}{2}\omega
(\phi)\dot{\phi}^2+V(\phi)-\frac{3H\dot{\Omega}(\phi)}{\kappa^2}\,
,
\end{equation}
\begin{equation}\label{eq1appendix1}
-\frac{2 \Omega (\phi)\dot{H}}{\kappa^2}=\omega
(\phi)\dot{\phi}^2-\frac{H\dot{\Omega}(\phi)}{\kappa^2}+\frac{\ddot{\phi}\Omega
'(\phi)+\Omega'' (\phi)\dot{\phi}^2}{\kappa^2}\, ,
\end{equation}
\begin{equation}\label{eq1appendix2}
\ddot{\phi}+3 H\dot{\phi}+\frac{1}{\omega
(\phi)}\left(V'(\phi)-\frac{R \Omega'(\phi)}{2\kappa^2} \right)\,
,
\end{equation}
where the ``prime'' indicates differentiation with respect to the
scalar field. Apparently, the simultaneous presence of the
functions $\omega (\phi)$, $\Omega (\phi)$ and $V(\phi)$ makes
simply impossible to analytically study the inflationary dynamics
in the $\phi$-Jordan frame for the generalized $R^p$ inflation.
Thus one must rely on the Einstein frame calculation and the
conformal equivalence of the frames in order to study analytically
the the generalized $R^p$ inflationary theory.


\begin{thebibliography}{99}



\bibitem{CMB-S4:2016ple}
K.~N.~Abazajian \textit{et al.} [CMB-S4],
[arXiv:1610.02743 [astro-ph.CO]].



\bibitem{SimonsObservatory:2019qwx}
M.~H.~Abitbol \textit{et al.} [Simons Observatory],
Bull. Am. Astron. Soc. \textbf{51} (2019), 147 [arXiv:1907.08284
[astro-ph.IM]].




\bibitem{Baker:2019nia}
J.~Baker, J.~Bellovary, P.~L.~Bender, E.~Berti, R.~Caldwell,
J.~Camp, J.~W.~Conklin, N.~Cornish, C.~Cutler and R.~DeRosa,
\textit{et al.}
[arXiv:1907.06482 [astro-ph.IM]].

\bibitem{Smith:2019wny}
T.~L.~Smith and R.~Caldwell,
Phys. Rev. D \textbf{100} (2019) no.10, 104055
doi:10.1103/PhysRevD.100.104055 [arXiv:1908.00546 [astro-ph.CO]].

\bibitem{Seto:2001qf}
N.~Seto, S.~Kawamura and T.~Nakamura,
Phys. Rev. Lett. \textbf{87} (2001), 221103
doi:10.1103/PhysRevLett.87.221103 [arXiv:astro-ph/0108011
[astro-ph]].

\bibitem{Kawamura:2020pcg}
S.~Kawamura, M.~Ando, N.~Seto, S.~Sato, M.~Musha, I.~Kawano,
J.~Yokoyama, T.~Tanaka, K.~Ioka and T.~Akutsu, \textit{et al.}
PTEP \textbf{2021} (2021) no.5, 05A105 doi:10.1093/ptep/ptab019
[arXiv:2006.13545 [gr-qc]].


\bibitem{Crowder:2005nr}
J.~Crowder and N.~J.~Cornish,
Phys. Rev. D \textbf{72} (2005), 083005
doi:10.1103/PhysRevD.72.083005 [arXiv:gr-qc/0506015 [gr-qc]].

\bibitem{Smith:2016jqs}
T.~L.~Smith and R.~Caldwell,
Phys. Rev. D \textbf{95} (2017) no.4, 044036
doi:10.1103/PhysRevD.95.044036 [arXiv:1609.05901 [gr-qc]].

\bibitem{Weltman:2018zrl}
A.~Weltman, P.~Bull, S.~Camera, K.~Kelley, H.~Padmanabhan,
J.~Pritchard, A.~Raccanelli, S.~Riemer-S\o{}rensen, L.~Shao and
S.~Andrianomena, \textit{et al.}
Publ. Astron. Soc. Austral. \textbf{37} (2020), e002
doi:10.1017/pasa.2019.42 [arXiv:1810.02680 [astro-ph.CO]].

\bibitem{NANOGrav:2020bcs}
Z.~Arzoumanian \textit{et al.} [NANOGrav],
Astrophys. J. Lett. \textbf{905} (2020) no.2, L34
doi:10.3847/2041-8213/abd401 [arXiv:2009.04496 [astro-ph.HE]].

\bibitem{NANOGrav:2020spf}
N.~S.~Pol \textit{et al.} [NANOGrav],
Astrophys. J. Lett. \textbf{911} (2021) no.2, L34
doi:10.3847/2041-8213/abf2c9 [arXiv:2010.11950 [astro-ph.HE]].





\bibitem{inflation1}
 A.~D.~Linde,
 Lect.\ Notes Phys.\ {\bf 738} (2008) 1
 [arXiv:0705.0164 [hep-th]].

\bibitem{inflation2} D.~S.~Gorbunov and V.~A.~Rubakov,
``Introduction to the theory of the early universe: Cosmological
perturbations and inflationary theory,'' Hackensack, USA: World
Scientific (2011) 489 p;
%


\bibitem{inflation3}A.~Linde,
arXiv:1402.0526 [hep-th];


\bibitem{inflation4}D.~H.~Lyth and A.~Riotto,
Phys.\ Rept.\  {\bf 314} (1999) 1 [hep-ph/9807278].


\bibitem{bounce0}
R.~H.~Brandenberger,
[arXiv:1206.4196 [astro-ph.CO]].


\bibitem{bounce1} M.~Novello and S.~E.~P.~Bergliaffa,
  Phys.\ Rept.\  {\bf 463} (2008) 127
  doi:10.1016/j.physrep.2008.04.006
  [arXiv:0802.1634 [astro-ph]].


\bibitem{bounce2} C.~Li, R.~H.~Brandenberger and Y.~K.~E.~Cheung,
  Phys.\ Rev.\ D {\bf 90} (2014) no.12,  123535
  doi:10.1103/PhysRevD.90.123535
  [arXiv:1403.5625 [gr-qc]].


\bibitem{bounce3} J.~de Haro and Y.~F.~Cai,
  Gen.\ Rel.\ Grav.\  {\bf 47} (2015) no.8,  95
  [arXiv:1502.03230 [gr-qc]].

\bibitem{bounce4} Y.~F.~Cai, E.~McDonough, F.~Duplessis and R.~H.~Brandenberger,
  JCAP {\bf 1310} (2013) 024
  doi:10.1088/1475-7516/2013/10/024
  [arXiv:1305.5259 [hep-th]].;

S.~Nojiri, S.~D.~Odintsov and V.~K.~Oikonomou,
  Phys.\ Rev.\ D {\bf 93} (2016) no.8,  084050
  doi:10.1103/PhysRevD.93.084050
  [arXiv:1601.04112 [gr-qc]].

\bibitem{bounce5} S.~D.~Odintsov and V.~K.~Oikonomou,
  arXiv:1512.04787 [gr-qc].

\bibitem{bounce6} V.~K.~Oikonomou,
  Phys.\ Rev.\ D {\bf 92} (2015) no.12,  124027
  doi:10.1103/PhysRevD.92.124027
  [arXiv:1509.05827 [gr-qc]].



\bibitem{Baumann:2009ds}
D.~Baumann,
doi:10.1142/9789814327183\_0010 [arXiv:0907.5424 [hep-th]].




\bibitem{reviews1}
 S.~Nojiri, S.~D.~Odintsov and V.~K.~Oikonomou,
  Phys.\ Rept.\  {\bf 692} (2017) 1
  [arXiv:1705.11098 [gr-qc]].

\bibitem{reviews2}


 S. Capozziello, M. De Laurentis,
   Phys.\ Rept.\  {\bf 509}, 167 (2011);\\
V. Faraoni and S. Capozziello, The landscape beyond Einstein
gravity, in Beyond Einstein Gravity 828 (Springer, Dordrecht,
2010), Vol. 170, pp.59-106.



   \bibitem{reviews3}

S. Nojiri, S.D. Odintsov,
   Phys.\ Rept.\  {\bf 505}, 59 (2011);

\bibitem{alpha1} R.~Kallosh and A.~Linde,
  JCAP {\bf 1307} (2013) 002
  [arXiv:1306.5220 [hep-th]].




\bibitem{alpha2} S.~Ferrara, R.~Kallosh, A.~Linde and M.~Porrati,
  Phys.\ Rev.\ D {\bf 88} (2013) no.8,  085038
  [arXiv:1307.7696 [hep-th]].


\bibitem{alpha3}R.~Kallosh, A.~Linde and D.~Roest,
  JHEP {\bf 1311} (2013) 198
  [arXiv:1311.0472 [hep-th]].






\bibitem{alpha4}

A.~Linde,
JCAP \textbf{05} (2015), 003
[arXiv:1504.00663 [hep-th]].






\bibitem{alpha5} S.~Cecotti and R.~Kallosh,
  JHEP {\bf 1405} (2014) 114
  [arXiv:1403.2932 [hep-th]].



\bibitem{alpha6} J.~J.~M.~Carrasco, R.~Kallosh and A.~Linde,
  JHEP {\bf 1510} (2015) 147
  [arXiv:1506.01708 [hep-th]].



\bibitem{alpha7}   J.~J.~M.~Carrasco, R.~Kallosh, A.~Linde and D.~Roest,
Phys. Rev. D \textbf{92} (2015) no.4, 041301
doi:10.1103/PhysRevD.92.041301 [arXiv:1504.05557 [hep-th]].


\bibitem{alpha7a}

R.~Kallosh, A.~Linde and D.~Roest,
Phys. Rev. Lett. \textbf{112} (2014) no.1, 011303
doi:10.1103/PhysRevLett.112.011303 [arXiv:1310.3950 [hep-th]].


\bibitem{alpha8} D.~Roest and M.~Scalisi,
  Phys.\ Rev.\ D {\bf 92} (2015) 043525
  doi:10.1103/PhysRevD.92.043525
  [arXiv:1503.07909 [hep-th]].




\bibitem{alpha9}  R.~Kallosh, A.~Linde and D.~Roest,
  JHEP {\bf 1408} (2014) 052
  doi:10.1007/JHEP08(2014)052
  [arXiv:1405.3646 [hep-th]].

\bibitem{alpha10} J.~Ellis, D.~V.~Nanopoulos and K.~A.~Olive,
  JCAP {\bf 1310} (2013) 009
  [arXiv:1307.3537 [hep-th]].

\bibitem{alpha11} Y.~F.~Cai, J.~O.~Gong and S.~Pi,
  Phys.\ Lett.\ B {\bf 738} (2014) 20
  doi:10.1016/j.physletb.2014.09.009
  [arXiv:1404.2560 [hep-th]].

\bibitem{alpha12}
Z.~Yi and Y.~Gong,
Phys. Rev. D \textbf{94} (2016) no.10, 103527
doi:10.1103/PhysRevD.94.103527 [arXiv:1608.05922 [gr-qc]].



\bibitem{alpha13}

Y.~Akrami, R.~Kallosh, A.~Linde and V.~Vardanyan,
JCAP \textbf{06} (2018), 041 doi:10.1088/1475-7516/2018/06/041
[arXiv:1712.09693 [hep-th]].



\bibitem{alpha14}
S.~Qummer, A.~Jawad and M.~Younas,
Int. J. Mod. Phys. D \textbf{29} (2020) no.16, 2050117
doi:10.1142/S0218271820501175



\bibitem{alpha15}
Q.~Fei, Z.~Yi and Y.~Yang,
Universe \textbf{6} (2020) no.11, 213 doi:10.3390/universe6110213
[arXiv:2009.14819 [gr-qc]].



\bibitem{alpha16}
A.~D.~Kanfon, F.~Mavoa and S.~M.~J.~Houndjo,
Astrophys. Space Sci. \textbf{365} (2020) no.6, 97
doi:10.1007/s10509-020-03813-6


\bibitem{alpha17}
I.~Antoniadis, A.~Karam, A.~Lykkas, T.~Pappas and K.~Tamvakis,
PoS \textbf{CORFU2019} (2020), 073 doi:10.22323/1.376.0073
[arXiv:1912.12757 [gr-qc]].


\bibitem{alpha18}
C.~Garc\'\i{}a-Garc\'\i{}a, P.~Ru\'\i{}z-Lapuente, D.~Alonso and
M.~Zumalac\'arregui,
JCAP \textbf{07} (2019), 025 doi:10.1088/1475-7516/2019/07/025
[arXiv:1905.03753 [astro-ph.CO]].



\bibitem{alpha19}
F.~X.~Linares Cede\~no, A.~Montiel, J.~C.~Hidalgo and G.~Germ\'an,
JCAP \textbf{08} (2019), 002 doi:10.1088/1475-7516/2019/08/002
[arXiv:1905.00834 [gr-qc]].



\bibitem{alpha20}
S.~Karamitsos,
JCAP \textbf{09} (2019), 022 doi:10.1088/1475-7516/2019/09/022
[arXiv:1903.03707 [hep-th]].


\bibitem{alpha21}
D.~D.~Canko, I.~D.~Gialamas and G.~P.~Kodaxis,
Eur. Phys. J. C \textbf{80} (2020) no.5, 458
doi:10.1140/epjc/s10052-020-8025-4 [arXiv:1901.06296 [hep-th]].


\bibitem{alpha22}
T.~Miranda, C.~Escamilla-Rivera, O.~F.~Piattella and J.~C.~Fabris,
JCAP \textbf{05} (2019), 028 doi:10.1088/1475-7516/2019/05/028
[arXiv:1812.01287 [gr-qc]].


\bibitem{alpha23}
A.~Karam, T.~Pappas and K.~Tamvakis,
JCAP \textbf{02} (2019), 006 doi:10.1088/1475-7516/2019/02/006
[arXiv:1810.12884 [gr-qc]].


\bibitem{alpha24}
K.~Nozari and N.~Rashidi,
Astrophys. J. \textbf{863} (2018) no.2, 133
doi:10.3847/1538-4357/aad18e [arXiv:1808.05363 [astro-ph.CO]].


\bibitem{alpha25}
C.~Garc\'\i{}a-Garc\'\i{}a, E.~V.~Linder, P.~Ru\'\i{}z-Lapuente
and M.~Zumalac\'arregui,
JCAP \textbf{08} (2018), 022 doi:10.1088/1475-7516/2018/08/022
[arXiv:1803.00661 [astro-ph.CO]].



\bibitem{alpha26}
N.~Rashidi and K.~Nozari,
Int. J. Mod. Phys. D \textbf{27} (2018) no.07, 1850076
doi:10.1142/S0218271818500761 [arXiv:1802.09185 [astro-ph.CO]].



\bibitem{alpha27}
Q.~Gao, Y.~Gong and Q.~Fei,
JCAP \textbf{05} (2018), 005 doi:10.1088/1475-7516/2018/05/005
[arXiv:1801.09208 [gr-qc]].


\bibitem{alpha28}
K.~Dimopoulos, L.~Donaldson Wood and C.~Owen,
Phys. Rev. D \textbf{97} (2018) no.6, 063525
doi:10.1103/PhysRevD.97.063525 [arXiv:1712.01760 [astro-ph.CO]].



\bibitem{alpha29}
T.~Miranda, J.~C.~Fabris and O.~F.~Piattella,
JCAP \textbf{09} (2017), 041 doi:10.1088/1475-7516/2017/09/041
[arXiv:1707.06457 [gr-qc]].



\bibitem{alpha30}
A.~Karam, T.~Pappas and K.~Tamvakis,
Phys. Rev. D \textbf{96} (2017) no.6, 064036
doi:10.1103/PhysRevD.96.064036 [arXiv:1707.00984 [gr-qc]].



\bibitem{alpha31}
K.~Nozari and N.~Rashidi,
Phys. Rev. D \textbf{95} (2017) no.12, 123518
doi:10.1103/PhysRevD.95.123518 [arXiv:1705.02617 [astro-ph.CO]].



\bibitem{alpha32}
Q.~Gao and Y.~Gong,
Eur. Phys. J. Plus \textbf{133} (2018) no.11, 491
doi:10.1140/epjp/i2018-12324-3 [arXiv:1703.02220 [gr-qc]].


\bibitem{alpha33}
C.~Q.~Geng, C.~C.~Lee and Y.~P.~Wu,
Eur. Phys. J. C \textbf{77} (2017) no.3, 162
doi:10.1140/epjc/s10052-017-4720-1 [arXiv:1512.04019
[astro-ph.CO]].


\bibitem{alpha34}
S.~D.~Odintsov and V.~K.~Oikonomou,
Phys. Lett. B \textbf{807} (2020), 135576
doi:10.1016/j.physletb.2020.135576 [arXiv:2005.12804 [gr-qc]].



\bibitem{alpha35}
S.~D.~Odintsov and V.~K.~Oikonomou,
Phys. Rev. D \textbf{94} (2016) no.12, 124026
doi:10.1103/PhysRevD.94.124026 [arXiv:1612.01126 [gr-qc]].



\bibitem{alpha36}
S.~D.~Odintsov and V.~K.~Oikonomou,
Class. Quant. Grav. \textbf{34} (2017) no.10, 105009
doi:10.1088/1361-6382/aa69a8 [arXiv:1611.00738 [gr-qc]].


\bibitem{alpha37}

L.~J\"arv, A.~Karam, A.~Kozak, A.~Lykkas, A.~Racioppi and M.~Saal,
Phys. Rev. D \textbf{102} (2020) no.4, 044029
doi:10.1103/PhysRevD.102.044029 [arXiv:2005.14571 [gr-qc]].


\bibitem{Ivanov:2021ily}
V.~R.~Ivanov and S.~Y.~Vernov,
Eur. Phys. J. C \textbf{81} (2021) no.11, 985
doi:10.1140/epjc/s10052-021-09792-4 [arXiv:2108.10276 [gr-qc]].



\bibitem{Motohashi:2014tra}
H.~Motohashi,
Phys. Rev. D \textbf{91} (2015), 064016
doi:10.1103/PhysRevD.91.064016 [arXiv:1411.2972 [astro-ph.CO]].


\bibitem{Renzi:2019ewp}
F.~Renzi, M.~Shokri and A.~Melchiorri,
Phys. Dark Univ. \textbf{27} (2020), 100450
doi:10.1016/j.dark.2019.100450 [arXiv:1909.08014 [astro-ph.CO]].




  \bibitem{kaizer}N.~Deruelle and M.~Sasaki,
  Springer Proc.\ Phys.\  {\bf 137} (2011) 247
  [arXiv:1007.3563 [gr-qc]].;\\
M.~Li,
  Phys.\ Lett.\ B {\bf 736} (2014) 488
   Erratum: [Phys.\ Lett.\ B {\bf 747} (2015) 562]
  [arXiv:1405.0211 [hep-th]].;\\
J.~O.~Gong, J.~c.~Hwang, W.~I.~Park, M.~Sasaki and Y.~S.~Song,
  JCAP {\bf 1109} (2011) 023
  doi:10.1088/1475-7516/2011/09/023
  [arXiv:1107.1840 [gr-qc]].


\bibitem{newsergei} D.~I.~Kaiser, astro-ph/9507048; G. Domenech, M. Sasaki, arXiv:1602.06332;
  D.~J.~Brooker, S.~D.~Odintsov and R.~P.~Woodard,
  Nucl.\ Phys.\ B {\bf 911} (2016) 318
  [arXiv:1606.05879 [gr-qc]].; D.~I.~Kaiser,
  Phys.\ Rev.\ D {\bf 52} (1995) 4295
  [astro-ph/9408044].



\bibitem{Stabile:2013eha}
A.~Stabile, A.~Stabile and S.~Capozziello,
Phys. Rev. D \textbf{88} (2013) no.12, 124011
doi:10.1103/PhysRevD.88.124011 [arXiv:1310.7097 [gr-qc]].



\bibitem{Sotani:2017pfj}
H.~Sotani and K.~D.~Kokkotas,
Phys. Rev. D \textbf{95} (2017) no.4, 044032
doi:10.1103/PhysRevD.95.044032 [arXiv:1702.00874 [gr-qc]].



\bibitem{Astashenok:2021peo}
A.~V.~Astashenok, S.~Capozziello, S.~D.~Odintsov and
V.~K.~Oikonomou,
Phys. Lett. B \textbf{816} (2021), 136222
doi:10.1016/j.physletb.2021.136222 [arXiv:2103.04144 [gr-qc]].


\bibitem{Feola:2019zqg}
P.~Feola, X.~J.~Forteza, S.~Capozziello, R.~Cianci and S.~Vignolo,
Phys. Rev. D \textbf{101} (2020) no.4, 044037
doi:10.1103/PhysRevD.101.044037 [arXiv:1909.08847 [astro-ph.HE]].


\bibitem{Staykov:2014mwa}
K.~V.~Staykov, D.~D.~Doneva, S.~S.~Yazadjiev and K.~D.~Kokkotas,
JCAP \textbf{10} (2014), 006 doi:10.1088/1475-7516/2014/10/006
[arXiv:1407.2180 [gr-qc]].












\bibitem{starob1} A.~A.~Starobinsky,
  Phys.\ Lett.\ B {\bf 91} (1980) 99.
  doi:10.1016/0370-2693(80)90670-X



\bibitem{starob2}   J.~D.~Barrow and S.~Cotsakis,
  Phys.\ Lett.\ B {\bf 214} (1988) 515.
  doi:10.1016/0370-2693(88)90110-4







\bibitem{higgs} F.~L.~Bezrukov and M.~Shaposhnikov,
  Phys.\ Lett.\ B {\bf 659} (2008) 703
  doi:10.1016/j.physletb.2007.11.072
  [arXiv:0710.3755 [hep-th]].






\bibitem{Odintsov:2021nqa}
S.~D.~Odintsov and V.~K.~Oikonomou,
Annals Phys. \textbf{440} (2022), 168839
doi:10.1016/j.aop.2022.168839 [arXiv:2104.01982 [gr-qc]].



\bibitem{Oikonomou:2021iid}
V.~K.~Oikonomou,
Class. Quant. Grav. \textbf{38} (2021) no.17, 175005
doi:10.1088/1361-6382/ac161c [arXiv:2107.12430 [gr-qc]].





\bibitem{Planck:2018jri}
Y.~Akrami \textit{et al.} [Planck],
Astron. Astrophys. \textbf{641} (2020), A10
doi:10.1051/0004-6361/201833887 [arXiv:1807.06211 [astro-ph.CO]].



\bibitem{Hwang:2005hb}
J.~c.~Hwang and H.~Noh,
Phys. Rev. D \textbf{71} (2005), 063536
doi:10.1103/PhysRevD.71.063536 [arXiv:gr-qc/0412126 [gr-qc]].



\bibitem{Odintsov:2020thl}
S.~D.~Odintsov and V.~K.~Oikonomou,
Phys. Lett. B \textbf{807} (2020), 135576
doi:10.1016/j.physletb.2020.135576 [arXiv:2005.12804 [gr-qc]].



\end{thebibliography}
\end{document}